\documentclass[12pt,twoside]{article}
\usepackage{amsmath,mathrsfs,bm,amssymb,color}

\usepackage[dvips]{graphicx}
\usepackage{float}
\usepackage{latexsym}
\catcode`\@=11
\def\newsymbol#1#2#3#4#5{\let\next@\relax%
 \ifnum#2=\@ne\else%
 \ifnum#2=\tw@\let\next@\msyfam@\fi\fi%
 \mathchardef#1="#3\next@#4#5}
\def\mathhexbox@#1#2#3{\relax%
 \ifmmode\mathpalette{} {\m@th\mnnathchar"#1#2#3}
 \else\leavevmode\hbox{$\m@th\mathchar"#1#2#3$}\fi}
\font\tenmsy=msbm10
\font\sevenmsy=msbm7
\font\fivemsy=msbm5
\newfam\msyfam
\textfont\msyfam=\tenmsy
\scriptfont\msyfam=\sevenmsy
\scriptscriptfont\msyfam=\fivemsy
\edef\msyfam@{\hexnumber@\msyfam}
\def\mathbb #1{\fam\msyfam\relax#1}
\catcode`\@=\active

\newtheorem{theorem}{Theorem}[section]
\newtheorem{proposition}[theorem]{Proposition}
\newtheorem{lemma}[theorem]{Lemma}
\newtheorem{corollary}[theorem]{Corollary}
\newtheorem{definition}[theorem]{Definition}

\newtheorem{remark}[theorem]{Remark}

\newcommand{\bd}[1]{\begin{definition}\label{#1}}
\newcommand{\ed}{\end{definition}}

\newcommand{\vp}{\varphi}

\newcommand{\bl}[1]{\begin{lemma}\label{#1}}
\newcommand{\el}{\end{lemma}}
\newcommand{\bc}[1]{\begin{corollary}\label{#1}}
\newcommand{\ec}{\end{corollary}}
\newcommand{\bt}[1]{\begin{theorem}\label{#1}}
\newcommand{\et}{\end{theorem}}
\newcommand{\bp}[1]{\begin{proposition}\label{#1}}
\newcommand{\ep}{\end{proposition}}
\newcommand{\br}[1]{\begin{remark}\label{#1}}
\newcommand{\er}{\end{remark}}
\newcommand{\zz}{{\mathbb  Z}}
\renewcommand{\tt}{{\mathbb  T}}

\newcommand{\eq}[1]{\begin{equation}\label{#1}}
\newcommand{\en}{\end{equation}}
\newcommand{\eqn}{\begin{eqnarray*}}
\newcommand{\enn}{\end{eqnarray*}}
\newcommand{\eqnn}{\begin{eqnarray}}
\newcommand{\ennn}{\end{eqnarray}}

\newcommand{\qed}{\hfill {\bf qed}\par\medskip}
\newcommand{\bi}{\begin{description}}
\newcommand{\ei}{\end{description} }

\newcommand{\RR}{{\mathbb  R}}
\newcommand{\jjj}{\sum_{j=1}^d}

\newcommand{\kak}[1]{(\ref{#1})}

\newcommand{\ms}{\mathscr }

\newcommand{\ov}[1]{\overline{#1}}
\newcommand{\f}{^{-1}}
\newcommand{\lk}{\left(}
\newcommand{\rk}{\right)}
\newcommand{\lkk}{\left\{}
\newcommand{\rkk}{\right\}}

\def\bbbone{{\mathchoice {\rm 1\mskip-4mu l} {\rm 1\mskip-4mu l}
{\rm 1\mskip-4.5mu l} {\rm 1\mskip-5mu l}}}
\def\one{\bbbone}

\newcommand{\s}{\sigma}
\newcommand{\hhh}{{\ms H}}
\newcommand{\LR}{L^2(\tt^d)}

\newcommand{\lr}[1]{\left\langle#1\right\rangle_\vp}
\newcommand{\hz}{H_0}

\makeatletter
\@addtoreset{equation}{section}
\makeatother

\title
{\sc Note on the spectrum of  discrete Schr\"odinger operators}

\author{
 Fumio Hiroshima\\
Faculty of Mathematics, Kyushu University\\
Fukuoka, 819-0395, Japan
\\ [0.3cm]
 Itaru Sasaki\\
Fiber-Nanotech Young Researcher Empowerment Center, \\
Shinshu University,
Matsumoto 390-8621, Japan
     \\[0.3cm]
Tomoyuki Shirai\\
Faculty of Mathematics, Kyushu Unicersity\\
Fukuoka, 819-0395, Japan
    \\and \\
 Akito Suzuki\\
Department of Mathematics, Faculty of Engineering, \\
Shinshu University, Nagano 380-8553, Japan
}

\bigskip\bigskip\bigskip

\date{}

\begin{document}
\maketitle
\abstract{%
The spectrum of discrete Schr\"odinger operator $L+V$ on the
$d$-dimensional lattice is considered, where $L$ denotes the discrete
Laplacian and $V$ a delta function with mass at a single point.
Eigenvalues of $L+V$ are specified and the absence of singular
continuous spectrum is proven.  In particular it is shown that an
embedded eigenvalue does appear for $d\geq5$ but does not for $1\leq
d\leq 4$.
}

\section{Introduction}
In this paper we are concerned with the spectrum of $d$-dimensional
discrete Schr\"odinger operators on square lattices.  Let
$\ell^2(\zz^d)$ be the set of $\ell^2$ sequences on the $d$-dimensional
lattice $\zz^d$.  We consider the spectral property of a bounded
self-adjoint operator defined on $\ell^2(\zz^d)$:
\begin{align}
L+V,
\end{align}
where the  $d$-dimensional discrete Laplacian $L$  is defined  by 
\begin{align}
L\psi(x)=\frac{1}{2d} \sum_{|x-y|=1}\psi(y)
\end{align}
and the  interaction $V$  by 
\begin{align}
V\psi(x)=v\delta_0(x)\psi(x). 
\end{align}
Here  $v>0$ is a non-negative coupling constant and $\delta_0(x)$ denotes the delta function with mass at 
$0\in\zz^d$, i.e., 
$\delta_0(x)=\lkk 
\begin{array}{ll}
1,&x=0\\
0,&x\not=0.
\end{array}
\right.
$
To study the spectrum of $L+V$ we form $L+V$ by the Fourier transformation.  
Let $\tt^d=[-\pi,\pi]^d$ be the $d$-dimensional torus, and 
$F:\ell^2(\zz^d)\to L^2(\tt^d)$  be the Fourier transformation  defined by 
$$(F\psi)(\theta)=\sum_{x\in\zz^d}\psi(n)e^{-ix\cdot\theta},$$
where $\theta=(\theta_1,...,\theta_d)\in\tt^d$. 
The inverse Fourier transformation  is then given by 
$$(F^{-1}\psi)(x)=\frac{1}{(2\pi)^d} \int _{\tt^d} \psi(\theta) e^{ix\cdot \theta} d\theta.$$
Hence $L+V$ is transformed to a self-adjoint operator on $\LR$:
\begin{align}\label{fourier}
{F(L+V)F^{-1}\psi(\theta)} 
= \lk\frac{1}{d}\sum_{j=1}^d \cos\theta_j\rk\psi(\theta) +
\frac{v}{(2\pi)^d}\int_{\tt^d}\psi(\theta) d\theta.
\end{align}
In what follows we denote the right-hand side of \kak{fourier} by $H=H(v)$, 
and we set 
 $H(0)=\hz$. 
Thus 
\begin{align}
H=g+v(\vp,\cdot)_{\LR}\vp,\quad \vp=(2\pi)^{-d/2}\one,
\end{align}
where $(\cdot,\cdot)_{\LR}$ denotes the scalar product on $\LR$, which is linear in the right-component and anti-linear  in the left-component, 
and 
$g$ is the multiplication by 
the real-valued function:
\begin{align}
g(\theta)=\frac{1}{d}\sum_{j=1}^d \cos \theta_j.
\end{align}
Hence 
$H$ can be realized as a rank-one perturbation of the discrete Laplacian $g$.
We study the spectrum of $H$. 
We denote the spectrum (resp. point spectrum, discrete spectrum,
absolutely continuous spectrum, singular continuous spectrum,essential
spectrum) of self-adjoint operator 
$T$ by 
$\s(T)$ (resp. $\s_{\rm p}(T), \s_{\rm d}(T), \s_{\rm ac}(T),\s_{\rm sc}(T),\s_{\rm ess}(H)$).

\section{Results}
In the continuous case the Schr\"odinger operator is defined by 
$H_S=-\Delta+vV$ in $L^2(\RR^d)$. 
Let $V\geq 0$ and $V\in L_{loc}^1(\RR^d)$. 
Let $N$ denote the number of strictly negative eigenvalues of $H_S$. 
It is known that $N\geq1$  for all values of $v>0$ 
for $d=1,2$ \cite{sim05}. However in the case of $d\geq 3$, 
by the Lieb-Thirring bound \cite{lie76}
$N\leq a\int|vV(x)|^{d/2}dx$ follows 
with some constant $a$ independent of $V$. In particular 
for sufficiently small $v$, it follows that $N=0$.
For the discrete case similar results to those of the continuous version 
may be   
expected. 
We summarize
the  result obtained in this paper  below. 
\bt{main}
The spectrum of $H$ is as follows:
\bi
\item[($\s_{\rm ac}(H)$ and $\s_{\rm ess}(H)$)]
 $\s_{\rm ac}(H)=\s_{\rm ess}(H)=[-1,1]$ for all $v\geq 0$ and $d\geq 1$.
\item[($\s_{\rm sc}(H)$)]
  $\s_{\rm sc}(H)=\emptyset$ for all $v\geq 0$ and $d\geq 1$. 
\item[($\s_{\rm p}(H)$)]\ 
\bi
\item[($d=1,2$)]
For each  $v>0$, 
there exists $E>1$ such that 
$
\s_{\rm p}(H)=\s_{\rm d}(H)=\{E\}$.
In particular $E=\sqrt{1+ v^2}$ in the case of $d=1$.
\item[($d=3,4$)] \ 
\bi
\item
[($v>v_c$)]
There exists $E>1$ such that 
$\s_{\rm p}(H)=\s_{\rm d}(H)=\{E\}$.
\item
[($v\leq v_c$)]
  $\s_{\rm p}(H)=\emptyset$. 
\ei
\item[($d\geq 5$)] \ 
\bi
\item
[($v>v_c$)]
There exists $E>1$ such that 
$\s_{\rm p}(H)=\s_{\rm d}(H)=\{E\}$.
\item[($v=v_c$)]
 $\s_{\rm p}(H)=\{1\}$.
 \item
[($v< v_c$)]
  $\s_{\rm p}(H)=\emptyset$. 
\ei
\ei
\ei
\et
We give the proof of Theorem \ref{main} in Section 3 below.  The
absolutely continuous spectrum $\s_{\rm ac}(H)$ and essential spectrum
$\s_{\rm ess}(H)$ are discussed in Section \ref{absolutelycontinuous},
eigenvalues $\s_{\rm p}(H) $ in Theorem \ref{eigenvalue} and Theorem
\ref{absenceofembeddedeigenvalue}, and singular continuous spectrum
$\s_{\rm sc}(H)$ in Theorem~\ref{absenceofsc}.

\section{Spectrum}
\subsection{Absolutely continuous spectrum and essential spectrum }
\label{absolutelycontinuous}
It is known and fundamental to show that $\s_{\rm ac}(H)=\s_{\rm ess}(H)=[-1,1]$. 
Note that $\s(\hz )=\s_{\rm ac}(\hz )=\s_{\rm ess}(H)=[-1,1]$ 
is purely absolutely continuous spectrum and purely essential spectrum. 
Since the  perturbation $v(\vp,\cdot)\vp$ is a rank-one operator, 
the essential spectrum leaves invariant. Then $\s_{\rm ess}(H)=[-1,1]$.   
Let $\hhh_{\rm ac}$ denote the absolutely continuous part of $H$. 
The self-adjoint operator $H$ is a rank-one perturbation of $g$. 
Then the wave operator $W_\pm=\lim_{t\to\pm\infty}e^{itH(v)}e^{-it \hz } $ exists and is complete, 
which implies that $\hz $ and $H(v)\lceil_{\hhh_{\rm ac}}$ are unitarily equivalent by 
$W_\pm^{-1}\hz W_\pm=H(v)\lceil_{\hhh_{\rm ac}}$. 
In particular $\s_{\rm ac}(H)=\s_{\rm ac}(\hz )=[-1,1]$ follows. 

\subsection{Eigenvalues}
\subsubsection{Absence of embedded eigenvalues in $[-1,1)$}
In this section we discuss eigenvalues  of $H$. 
Namely we study the eigenvalue problem $H\psi=E\psi$, i.e., 
\begin{align}
v(\vp, \psi)\vp =(E-g)\psi.
\end{align}
The key lemma is as follows. 
\bl{discretespectrum}
$E\in \s_{\rm p}(H)$ if and only if 
\begin{align}\label{key}
\frac{1}{E-g}\in L^2(\tt^d)\quad 
\mbox{and}\quad v=(2\pi)^d \lk 
\int_{\tt^d} \frac{1}{E-g(\theta)} d\theta\rk ^{-1}.
\end{align}
Furthermore when $E\in \s_{\rm p}(H)$, 
it follows that 
$$H\frac{1}{E-g}=E\frac{1}{E-g},$$
 i.e., 
$\frac{1}{E-g}$ is the eigenvector associated with $E$. 
In particular every eigenvalue is simple.  
\el
\begin{proof}
Suppose that $E\in \s_{\rm p}(H)$. 
Then 
$(E-g)\psi=v(\vp,\psi)\vp$. 
Since $\psi\in\LR$ and $(E-g)\psi$ is a constant, 
$E-g\not=0$ almost everywhere and 
$\psi=v(\vp,\psi)\vp/(E-g)$  follows. 
Thus $(E-g)^{-1}\in\LR$. 
Inserting $\psi=c (E-g)^{-1}$ with some constant $c$ on both sides of 
$(E-g)\psi=v(\vp,\psi)\vp$, 
we obtain the second  identity in \kak{key} 
and then the necessity  part follows.
\end{proof}
The sufficiency  part can be easily seen. 
We state the absence of embedded eigenvalues in the interval $[-1,1)$. 
This can be derived from \kak{key}.
We summarize it in the theorem below:
\bt{absenceofembeddedeigenvalue}
 $\s_{\rm p}(H)\cap [-1,1)=\emptyset$.
\et

Suppose that $-1\in \s_{\rm p}(H)$. 
Then there exists a non-zero vector $\psi$ such that 
$(\psi, (g+1)\psi)+v(\vp,\psi)^2=0$. Thus 
$(\psi, (g+1)\psi)=0$ and $|(\vp,\psi)|^2=0$ follow.  
However we see that $(\psi, (g+1)\psi)\not=0$,
 since $g$ has no eigenvalues (has purely absolutely continuous spectrum). 
Then it is enough to show 
$\s_{\rm p}(H)\cap (-1,1)=\emptyset$. 
We shall  check that $\frac{1}{E-g}\not\in \LR$ for $-1<E<1$. 
By a direct computation we have 
\begin{align*}
{\int_{\tt^d}\frac{1}{(E-g(\theta))^2}d\theta} 
=\int_{[-1-E,1-E]^d}\frac{1}{(\frac{1}{d}\jjj X_j)^2}\prod_{j=1}^d 
\frac{1}{\sqrt{1-(X_j+E)^2}} dX.
\end{align*}
Changing variables by $X_1=Z_1,\cdots, X_{d-1}=Z_{d-1}$ and $\jjj
X_j=Z$. Then we have 
\begin{align*}
{\int_{\tt^d}\frac{1}{(E-g(\theta))^2}d\theta}
&=
\int_{\ov{\Delta}}
\frac{1}{\frac{1}{d^2}Z^2}
\frac{1}{\sqrt{1-(Z-Z_1-\cdots-Z_{d-1}+E)^2}}\\
&\quad \times \lk
\prod_{j=1}^{d-1} \frac{1}{\sqrt{1-(Z_j+E)^2}}
\rk
 J dZ \prod_{j=1}^{d-1}dZ_j,
\end{align*}
where $J=|
\det  \frac{\partial(Z_1,...,Z_{d-1},Z)}{\partial(X_1,...,X_d)}|=1$ 
is a Jacobian and $\Delta$ denotes the inside of a $d$-dimensional convex polygon including the origin, since $-1<E<1$, 
and $\ov \Delta$ is the closure of $\Delta$. 
Then we can take a rectangle  such that 
 $[-\delta,\delta]^d\subset \Delta$ for sufficiently small $0<\delta$.  
We have 
 the lower bound 
\begin{align*}
\int_{\tt^d}\frac{1}{(E-g(\theta))^2}d\theta 
\geq {\rm const}\times (2\delta)^{d-1} 
d^2 \int_{-\delta}^\delta 
\frac{1}{Z^2}dZ
\end{align*}
and the right-hand side diverges. 
 Then the theorem follows from \kak{key}. 
 \qed
\subsubsection{Eigenvalues in $[1,\infty)$}
Operator $H$ is bounded by the bound $\|H\|\leq 1+v/(2\pi)^d$.  
Then by  Theorem \ref{absenceofembeddedeigenvalue} and $v>0$,  
 eigenvalues are included in the interval $[1,(2\pi)^d v+1]$ whenever they exist.
We define the critical value $v_c$ by 
\begin{align}
v_c=(2\pi)^{d}\lk 
\int_{\tt^d} \frac{1}{1-g(\theta)}d\theta\rk^{-1}\in[0,\infty)
\end{align}
with convention $\frac{1}{\infty}=0$.
 \bl{preparation}
\bi
\item[(1)]
The function $[1,\infty)\ni E\mapsto \int_{\tt^d}\frac{1}{E-g(\theta)} d\theta$ is continuously decreasing.
\item[(2)]
 $v_c=0$ for $d=1,2$ and 
 $v_c>0$ for $d\geq 3$.
\item[(3)]
$(E-g)^{-1}\in\LR$ for all $d\geq1$ and $E>1$. 
\item[(4)]
$(1-g)^{-1}\in\LR$ for $d\geq5$ and $(1-g)^{-1}\not\in\LR$ for $1\leq d\leq 4$.
\ei
\el
\begin{proof}
(1) and (3)  are  straightforward. 
In order to show (2) 
it  is enough to consider a  neighborhood $U$ of  points where 
the denominator $1-g(\theta)$ vanishes.  On $U$, approximately 
\eq{ap}
1-g(\theta)\approx \frac{1}{2d}\jjj \theta_j^2.\en
Then 
$$
\int_U \frac{1}{1-g(\theta)}d\theta
\approx
\int_U \frac{1}{ \frac{1}{2d}\jjj \theta_j^2}d\theta
\approx {\rm const}\times 
\int_{U'} \frac{r^{d-1}}{ r^2}d r.
$$ 
We have 
$\int_U \frac{1}{ \frac{1}{2d}\jjj \theta_j^2}d\theta<\infty$ for $d\geq 3$ and 
$\int_U \frac{1}{ \frac{1}{2d}\jjj \theta_j^2}d\theta=\infty$ for $d=1,2$.
 Then (2) follows. 
(4)  can be proven in a similar manner to  (2).
Since 
\begin{align*}
\int_U \frac{1}{(1-g(\theta))^2}d\theta
\approx
\int_U \frac{1}{ (\frac{1}{2d}\jjj \theta_j^2)^2}d\theta 
\approx {\rm const}\times 
\int_{U'} \frac{r^{d-1}}{ r^4}d r,
\end{align*}
we have $(1-g)^{-1}\in\LR$
 for $d\geq 5$ and 
$(1-g)^{-1}\not\in\LR$ for $d=1,2,3,4$.
\end{proof}

From this lemma we can immediately obtain results on eigenvalue problem of
\begin{align}
\label{eq}
v(\vp, \psi)\vp =(E-g)\psi.
\end{align}
\bt{eigenvalue}
\bi
\item[($d=1,2$)]
\kak{eq} has a unique solution $\psi=\frac{1}{E-g}$ up to a multiplicative constant and $E>1$ for each  $v>0$. 
In particular $E=\sqrt{1+ v^2}$ for $d=1$.
\item[($d=3,4$)]
\kak{eq} has the unique solution $\psi=\frac{1}{E-g}$ up to a multiplicative constant and $E>1$ for $v>v_c$ and no non-zero solution for $v\leq v_c$.
In particular $1$ is not eigenvalue for $H(v_c)$. 
\item[($d\geq5$)]
\kak{eq} has the unique solution $\psi=\frac{1}{E-g}$ up to a multiplicative constant and $E\geq 1$ for $v\geq v_c$ 
and  no non-zero solution for $v<v_c$.
In particular $E=1$ is eigenvalue for $H(v_c)$.
\ei
\et
\begin{proof}
In the case of $d=1,2$, \kak{key} is fulfilled for all $v>0$, and
$\frac{v}{2\pi}\int_{\tt^d}\frac{1}{E-g(\theta)}=1$ follows from
$H\frac{1}{E-g}=\frac{E}{E-g}$.  Thus $E=\sqrt{1+ v^2}$ for $d=1$.  In
the case of $d=3,4$, \kak{key} is fulfilled for $v> v_c$, but not for
$v=v_c$.  In the case of $d\geq 5$, \kak{key} is fulfilled for $v\geq
v_c$.
\end{proof}

\subsection{Absence of singular continuous spectrum}
Let $\lr{T}=(\vp,T\vp)$ be the expectation of $T$ with respect to $\vp$. 
We introduce three  subsets in $\RR$. 
Let 
\begin{align*}
X&=\lkk x\in\RR|{\rm Im} \lr{ (\hz -(x+i0))\f}>0\rkk\\
Y&=\lkk x\in\RR|{\lr{ (\hz -x)^{-2}}^{-1}}>0\rkk\\
Z&=\RR\setminus(X\cup Y).
\end{align*}
Note that ${\rm Im} \lr{ (\hz -(x+i\epsilon))\f}\leq \epsilon \lr{ (\hz
-x)^{-2}}$. Then $X,Y$ and $Z$ are mutually disjoint.  Let $\mu_v^{\rm
ac}$(resp. $\mu_v^{\rm sc}$ and $\mu_v^{\rm pp}$) be the spectral mesure
of the absolutely continuous spectral part of $H(v)$ (resp. singular
continuous part, point spectral part).  A key ingredient to prove the
absence of singular continuous spectrum of a self-adjoint operator with
rank-one perturbation is the result of \cite[Theorem 1(b) and Theorem
3]{sw86} and \cite{aro57}. We say that a measure $\eta$ is supported on
$A$ if $\eta(\RR\setminus A)=0$. \bp{sw86}
For any $v\not=0$, 
$\mu_v^{\rm ac}$ is supported on $X$, 
 $\mu_v^{\rm pp}$ is supported on $Y$ and 
 $\mu_v^{\rm sc}$ is supported on $Z$.
 In particular when $\RR\setminus X\cup Y$ is countable, 
 $\s_{\rm sc}(H)=\emptyset$ follows.
  \ep
\begin{proof}
The former result is due to \cite[Theorem 1(b) and Theorem 3]{sw86}.
Since any countable sets have $\mu_v^{\rm sc}$-zero measure, 
the latter statement also follows.
\end{proof}
 
\bt{absenceofsc}
$\s_{\rm sc}(H)=\emptyset$.
\et
\begin{proof}
We shall show that $\RR\setminus {X\cup Y}$ is countable. 
Let $E\in \s_{\rm p}(H)$. 
Then 
it is shown in \kak{key} that 
$\lr{(\hz -E)^{-2}}=
\int _{\tt^d}\frac{1}{(g(\theta)-E)^2}d\theta<\infty$. 
Then $E\in Y$. 
Let $x\in (-\infty,-1)\cup(1,\infty)$. It is clear  that 
$\lr{(\hz -E)^{-2}}<\infty$. Then 
\eq{1}
\s_{\rm p}(H)\cup(-\infty,-1)\cup(1,\infty)
\subset  Y.
\en
Let $x\in (-1,1)$.
Then $(x-g)^{-1}\not\in \LR$ follows from the proof of Theorem \ref{absenceofembeddedeigenvalue}.
We have 
 $$ 
  {\rm Im} \lr { (\hz -(x+i\epsilon))\f}
 =
 \int _{\tt^d}\frac{\epsilon}{(g(\theta)-x)^2+\epsilon^2}d\theta.$$
We can compute the the right-hand side above in the same way as in 
the proof of Theorem \ref{absenceofembeddedeigenvalue}:
\begin{align*}
 \int _{\tt^d}\frac{\epsilon}{(g(\theta)-x)^2+\epsilon^2}d\theta
 \geq 
 (2\delta)^{d-1} 
d^2 \int_{-\delta}^\delta dZ
\frac{\epsilon}{Z^2+\epsilon^2}.
\end{align*}
Then  the right-hand side above converges to $(2\delta)^{d-1}d^2 \pi >0$
as $\epsilon\downarrow 0$. Then 
\eq{2}(-1,1)\subset  X.\en
By \kak{1} and \kak{2},  $\RR\setminus X\cup Y\subset \{-1,1\}$,  
 the theorem follows from Proposition  \ref{sw86}.
\end{proof}

\section{Concluding remarks}
Our next  issue will be to consider the spectral properties of
discrete Schr\"odinger operators with the sum (possibly infinite sum) of
delta functions: \eq{n1} L+v\sum_{j=1}^n\delta_{a_j}\quad 1<n\leq\infty.
\en This is transformed to \eq{n} H=g+v \sum_{j=1}^n (\vp_j,\cdot)\vp_j
\en by the Fourier transformation, where $\vp_j=(2\pi)^{-d/2}e^{-i\theta
a_j}$. 
Note that
$$(\vp_i,\vp_j)=(2\pi)^{-d}\int_{\tt^d}e^{i(a_i-a_j)\theta}d\theta=\delta_{ij}.$$
When $n<\infty$, $H$ is a finite rank perturbation of $g$.  Then the
absolutely continuous spectrum and the essential spectrum of $H$ are
$[-1,1]$.  In this case the discrete spectrum is studied in e.g.,
\cite{hmo11} for $d=1$. See also \cite{dks05}.  The absence of singular
continuous spectrum of $H$ may be shown by an application of the Mourre
estimate \cite{mou80}.  In order to study eigenvalues we may need
further effort.

\section*{Acknowledgments}
We thank Yusuke  Higuchi for sending the problem to our attension and giving 
a lot of useful comments. 
We  also thank Hiroshi  Isozaki for a helpful comments. 
 FH  is financially supported by 
Grant-in-Aid for Science Research (B) 20340032
from JSPS. 
TS's work was supported in part 
by JSPS Grant-in-Aid for Scientific Research (B) 22340020.

\end{document}